%% file: OSS_myths_and_facts_japanese.tex
\documentclass[pdflatex, ja=standard]{bxjsreport}
\def\documentLanguage{Japanese}
\usepackage{OSS_myths_and_facts}

\begin{document}

\input{common/body}

\end{document}

%% file: common/body.tex
\input{common/metainfo}

\input{common/cover}
\maketitle
\tableofcontents

\input{\documentLanguage/chapter1}
    \input{\documentLanguage/myths}
\input{\documentLanguage/chapter2}
\input{\documentLanguage/chapter3}
    \input{\documentLanguage/chapter3_3}
    \input{\documentLanguage/chapter3_2} 
    \input{\documentLanguage/chapter3_6} 
    \input{\documentLanguage/chapter3_14} 
    \input{\documentLanguage/chapter3_9} 
    \input{\documentLanguage/chapter3_5} 
\input{\documentLanguage/chapter4}

\input{common/colphan}

%% file: common/metainfo.tex
\date{2024/04/01}

\ifthenelse{\equal{\documentLanguage}{English}}{
    \title{OSS Myths and Facts}
    \author{
        Yukako Iimura \and 
        Masanari Kondo \and 
        Kazushi Tomoto \and 
        Yasutaka Kamei \and 
        Naoyasu Ubayashi \and 
        Shinobu Saito
    }
}{
    \title{OSSの神話と真実}
    \author{
        飯村 結香子 \and 
        近藤 将成 \and 
        東本 知志 \and 
        亀井 靖高 \and 
        鵜林 尚靖 \and 
        斎藤 忍
    }
}

\hypersetup{
    pdftitle={OSS Myths and Facts},
    pdfauthor={Yukako Iimura, Masanari Kondo, Kazushi Tomoto, Yasutaka Kame, Naoyasu Ubayashi, Shinobu Saito}
}

\newcommand*{\CRtext}{\copyright 2024 Nippon Telegraph and Telephone Corporation, KYUSHU UNIVERSITY}

%% file: common/cover.tex
\enlargethispage{\paperwidth}
\thispagestyle{empty}
\noindent\hspace*{-1in}\hspace*{-\oddsidemargin}
\ifthenelse{\equal{\documentLanguage}{English}}{
    \includegraphics[keepaspectratio, width=0.8\paperwidth]{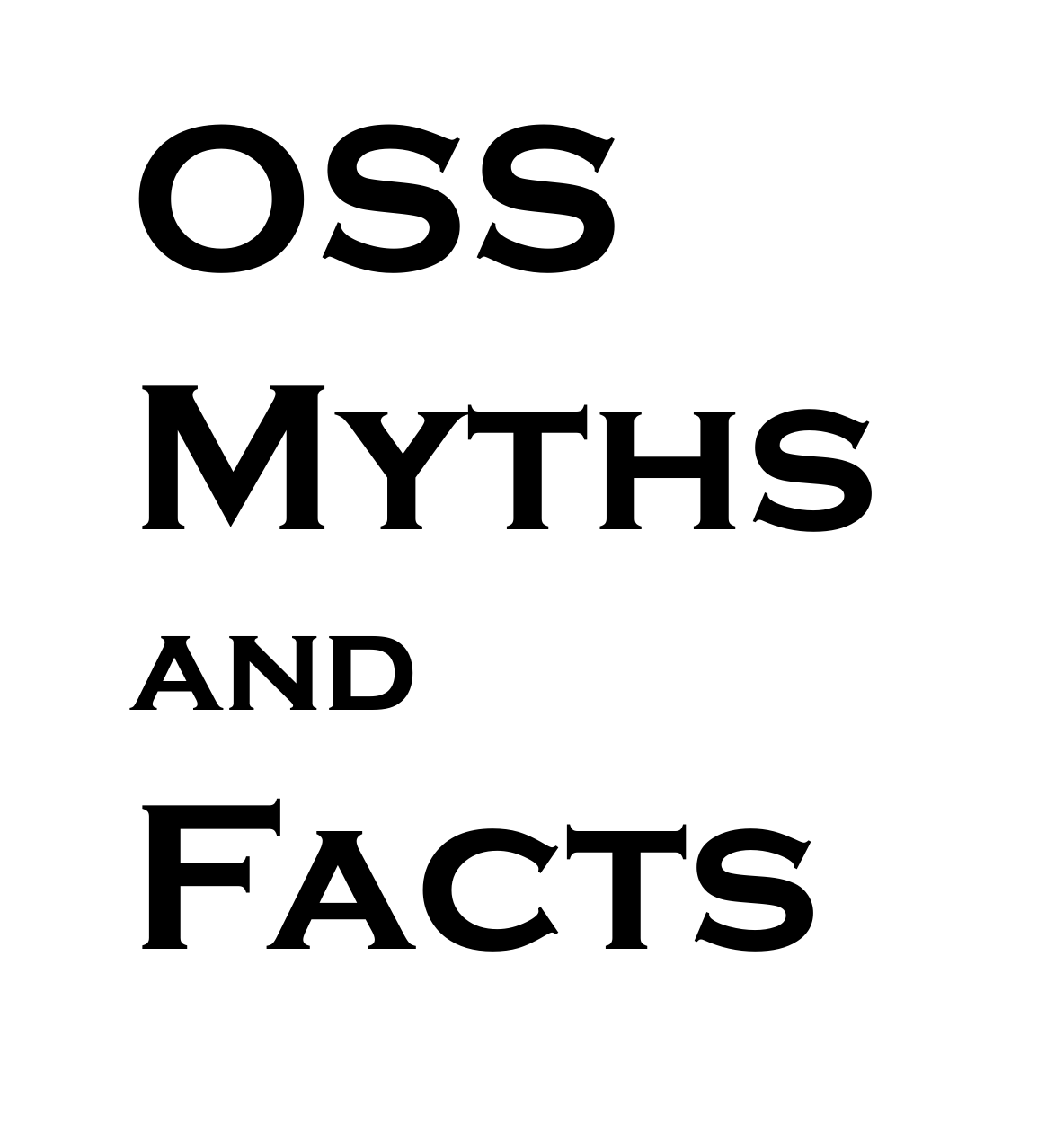}
}{}
\ifthenelse{\equal{\documentLanguage}{Japanese}}{
        \includegraphics[keepaspectratio, width=0.8\paperwidth]{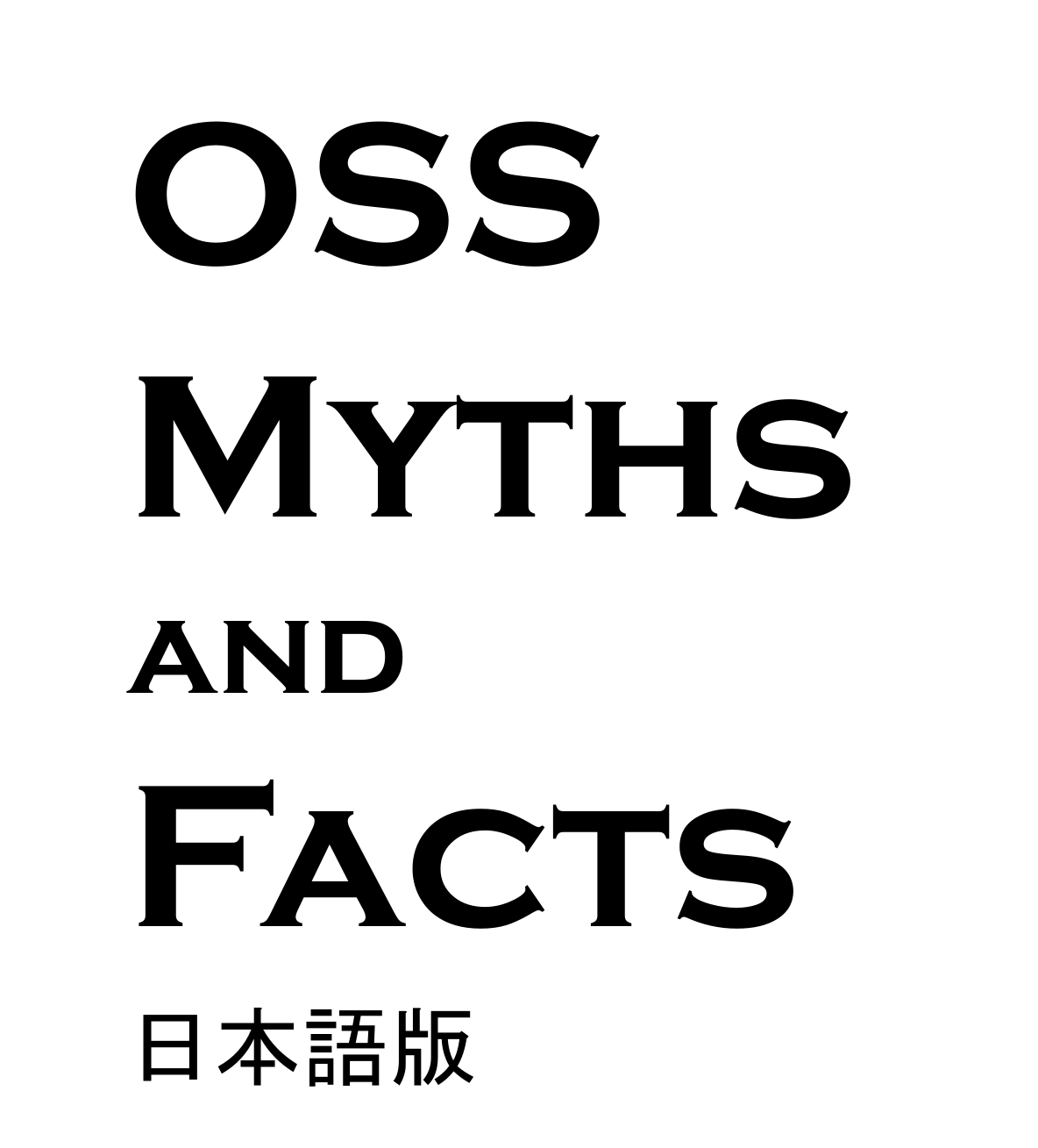}
}()
\pagebreak

%% file: English/chapter1.tex
\chapter{Executive Summary}

We have selected six myths about the OSS community and have tested
whether they are true or not.

The purpose of this report is to identify the lessons that can be
learned from the development style of the OSS community and the issues
that need to be addressed in order to achieve better Employee Experience
(EX) in software development within companies and organizations. The OSS
community has been led by a group of skilled developers known as
hackers. We have great respect for the engineers and activities of the
OSS community and aim to learn from them.

On the other hand, it is important to recognize that having high
expectations can sometimes result in misunderstandings. When there are
excessive expectations and concerns, misunderstandings (referred to as
myths) can arise, particularly when individuals who are not
practitioners rely on hearsay to understand the practices of
practitioners.

We selected the myths to be tested based on a literature review and
interviews. These myths are held by software development managers and
customers who are not direct participants in the OSS community. We
answered questions about each myth through: 1) Our own analysis of
repository data, 2) A literature survey of data analysis conducted by
previous studies, or 3) A combination of the two approaches.

\clearpage

%% file: English/myths.tex
\subsection*{Myth: Communication within the OSS is moderate.}

\noindent Question: Does the OSS community have long intervals in their
communication?\\
Fact: No matter the topic of discussion, developers communicate with
each other within a span of 4 hours for about half of all
communications.

\subsection*{Myth: OSS community never sleeps.}

\noindent Question: Are working hours distributed among developers in
the OSS community?\\
Fact: Developers tend to work during office hours in North America.

\subsection*{Myth: OSS community quickly halts their development.}

\noindent Question: How much of the OSS community will continue?\\
Fact: Four years after adoption, half of the OSS community activities
are still ongoing.

\subsection*{Myth: OSS community does not lose to crackers.}

\noindent Question: Does the OSS community take less time to resolve
vulnerabilities (security holes)?\\
Fact: A vulnerability resolution time of 3 months is not necessarily
short.

\subsection*{Myth: OSS community responds quickly to requests.}

\noindent Question: Is the resolution time for proposals, like bug
reports and enhancements, short?\\
Fact: Most bug reports and feature requests are resolved within two
weeks. However, at least a quarter of them take more than three months
to resolve, indicating a variation in resolution time.

\subsection*{Myth: OSS community participants are top-notch developers.}

\noindent Question: What roles do developers play in the OSS
Community?\\
Fact: In the OSS community, there is a wide range of roles available.

%% file: English/chapter2.tex
\chapter{Introduction}

\section{Employee Experience (EX) in Software Development}

We are conducting research and development with the goal of promoting
the well-being of individuals involved in software development and
enhancing the Employee Experience (EX). Our objective is to enable
software development to be carried out by anyone, at any time, and from
anywhere, by providing solutions that expand opportunities for diverse
individuals to thrive. Remote work, which has been promoted due to
COVID-19, is expected to broaden working conditions by expanding the
choice of work locations. Many companies cite objectives such as
``improving the well-being and health of employees,'' ``accommodating
individuals with commuting challenges,'' and ``retaining talented
individuals'' as reasons for implementing remote work. All of these
objectives contribute to creating a comfortable work environment for
employees. Workers have expressed various benefits of remote work,
including reduced commuting and travel time, increased free time, and
the ability to better balance work with childcare and parenting
responsibilities.

Traditionally, the workplace in software development, especially within
companies, has been restricted to a specific location. This is partly
because easy communication between clients and developers is believed to
lead to successful software development. Additionally, having a fixed
workplace facilitates information and worker management for companies.

\section{OSS as a predecessor}

``a world-class operating system could coalesce as if by magic out of
part-time hacking by several thousand developers scattered all over the
planet, connected only by the tenuous strands of the Internet'' from The
Cathedral and the Bazaar

Open Source Software(OSS) is a broad term that refers to software that
allows users to use, study, reuse, modify, extend, and redistribute its
source code for any purpose. It has gained significant attention for its
advanced development style, which involves frequent releases and
collaboration among contributors connected through the Internet. Much
valuable software is created through the cooperation of developers from
all over the world.

This fact has caught the attention of big tech, which also has something
to learn from it. As the OSS community grows, software developers and
management are becoming increasingly interested in the state of the OSS
community and product management.

\section{Is the OSS Community a Utopia?}

There are various opinions about the state of the OSS community and the
involvement of developers in projects. It is important to consider that
the opinions of clients and management are often subjective, influenced
by the origins of the OSS community and other factors. Practitioners
gain knowledge from their own experience, while the general population
receives and interprets empirical information, sometimes with
preconceived notions. Consequently, having overly high expectations can
lead to disappointment or excessive concern. For instance, the term
``agile'' is perceived differently by developers who associate it with
their own activities, and by management who view it as a means to an
end.

In this eBook, we will discuss six topics related to OSS and attempt to
distinguish between data-based facts and fiction (myths).

\section{Methodology}

\begin{enumerate}
\def\labelenumi{\arabic{enumi}.}
\item
  Topic Selection

  \begin{enumerate}
  \def\labelenumii{\arabic{enumii}.}
  \tightlist
  \item
    Gathering comments from software development clients and management.

    \begin{itemize}
    \tightlist
    \item
      Literature survey
    \item
      Interview
    \end{itemize}
  \item
    Organize interests and background
  \item
    3 categories and topics
  \end{enumerate}
\item
  Study each topic

  \begin{enumerate}
  \def\labelenumii{\arabic{enumii}.}
  \tightlist
  \item
    Primary study: Literature survey

    \begin{enumerate}
    \def\labelenumiii{\arabic{enumiii}.}
    \tightlist
    \item
      Literature survey
    \item
      Decision

      \begin{itemize}
      \tightlist
      \item
        If the topic has already been discussed in previous papers,
        provide a summary of the discussion
      \item
        Topics that have not yet been discussed or that can be further
        investigated are analyzed empirically
      \end{itemize}
    \end{enumerate}
  \item
    Secondary study: Data analysis

    \begin{itemize}
    \tightlist
    \item
      Main target dataset

      \begin{itemize}
      \tightlist
      \item
        Libraries.io open data
      \end{itemize}
    \item
      Analysis target extraction procedure

      \begin{enumerate}
      \def\labelenumiii{\arabic{enumiii}.}
      \tightlist
      \item
        list the products include in dataset
      \item
        Remove non-github repositories
      \item
        Remove fork repositories
      \item
        Filter for each analysis
      \item
        Random sampling
      \end{enumerate}
    \end{itemize}
  \end{enumerate}
\end{enumerate}

%% file: English/chapter3.tex
\chapter{Myths}

\begin{figure}[h]
\includegraphics[width=0.8\textwidth,height=0.8\textheight,keepaspectratio]{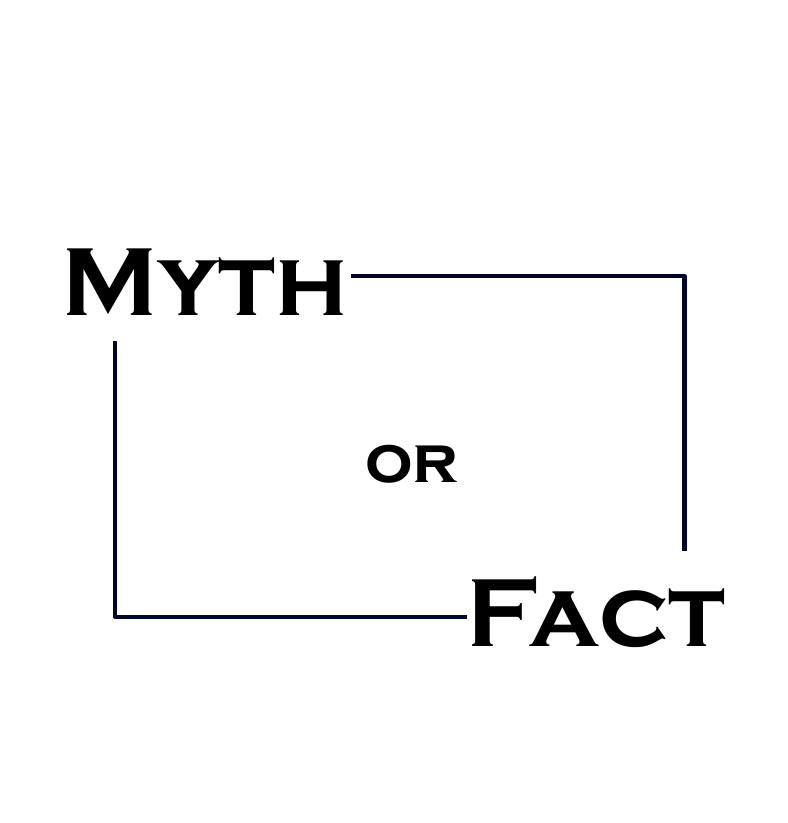}
\end{figure}

%% file: English/chapter3_3.tex
\section{Communication within the OSS is moderate.}

\subsection*{Question: Does the OSS community have long intervals in their communication?}

Software development is a collaborative process where communication
plays a key role. The pace of interactions between developers can
significantly impact a project's progress\autocite{cockburn}. In basic
software development, it's often considered best practice to physically
locate developers close to each other for this interaction pace.
However, this is not the case in the OSS community, where developers
don't typically share the same location or working hours. This situation
can make it challenging to initiate communication based on each team
member's preferred timing, leading to longer communication intervals,
but this is not justified by data. To investigate this myth, we analyze
the communication interval within the OSS community.

\subsection*{Fact: No matter the topic of discussion, developers communicate with each other within a span of 4 hours for about half of all communications.}

\begin{figure}[hbtp]
\centering
\includegraphics[width=0.9\textwidth,height=0.4\textheight,keepaspectratio]{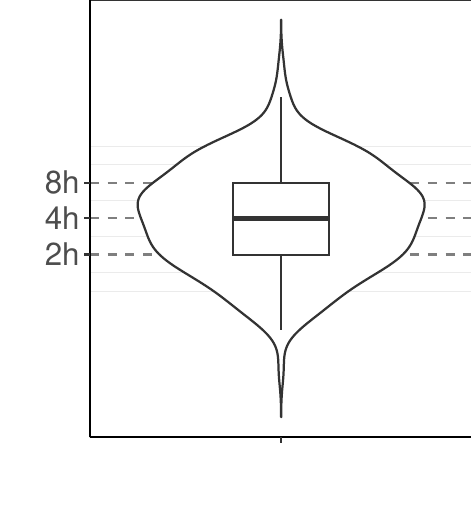}
\caption{The distribution of the time interval between a comment on an issue ticket}
\label{fig:comments_response}
\end{figure}

\begin{figure}[hbtp]
\centering
\includegraphics[width=0.9\textwidth,height=0.4\textheight,keepaspectratio]{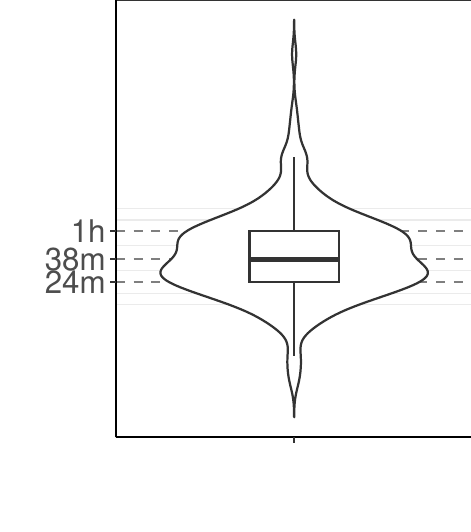}
\caption{The time interval distribution for a pull request}
\label{fig:PR_comment_time}
\end{figure}

We detail the communication time interval on GitHub for long-standing
OSSs \autocite{kondoCompsac23}. Figure \ref{fig:comments_response}
illustrates the distribution of the time interval between a comment on
an issue ticket and its subsequent reply. Figure
\ref{fig:PR_comment_time} displays the time interval distribution for a
pull request. The data suggest that the median time interval for issue
discussion is approximately 4 hours, while for pull requests, it's less
than 40 minutes. This suggests that communication within the OSS
community occurs at rapid intervals \autocite{kondoCompsac23}.

We infer that the brief response times are likely due to developers'
preference for synchronous communication, even in an asynchronous
environment (e.g., disparate locations and office hours). For instance,
Microsoft has reported that developers favor communication involving
numerous exchanges over a short duration\autocite{microsoft22}. However,
this could also indicate that developers feel pressured to respond
promptly. If so, they might feel obligated to answer at any time, even
when concentrating on other tasks, potentially negatively affecting
their well-being.

\subsection*{Insight:}

We observed that even the OSS community often operates in short time
intervals. In an era that accommodates diverse work styles, including
remote work, synchronous communication in corporate software development
may negatively impact developers' well-being. If software development
adopts more flexible communication, it could provide various benefits,
such as enabling developers to choose work hours that suit their
personal lifestyles.

\printbibliography[segment=\therefsegment,heading=subbibliography]

\clearpage

%% file: English/chapter3_2.tex
\section{OSS community never sleeps.}

\subsection*{Question: Are working hours distributed among developers in the OSS community?}

In basic distributed development, time and cultural differences can
sometimes cause certain developers to be ``overlooked'' or
``neglected,'' leading to confusion and discouragement. Given that the
open-source software (OSS) community includes developers worldwide
\autocite{githubGlobal}, a question arises: do developers in the OSS
community operate beyond time zone differences? If so, their working
hours should span across various time zones, making the OSS community
active round-the-clock. To examine this myth, we analyze the
distribution of developers' working hours in the OSS community.

\subsection*{Fact: Developers tend to work during office hours in North America.}

\begin{figure}[hbtp]
\centering
\includegraphics[width=0.9\textwidth,height=0.4\textheight,keepaspectratio]{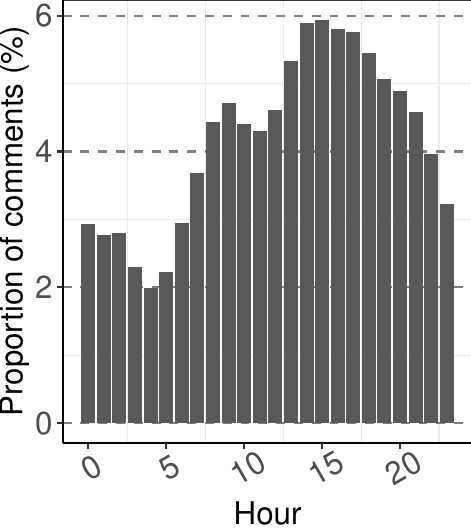}
\caption{The count of comments on an issue ticket}
\label{fig:comments_issue}
\end{figure}

\begin{figure}[hbtp]
\centering
\includegraphics[width=0.9\textwidth,height=0.4\textheight,keepaspectratio]{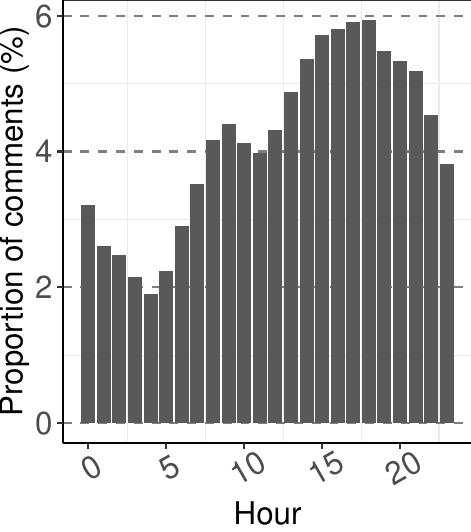}
\caption{The count of comments on pull requests}
\label{fig:comments_PR}
\end{figure}

Figure \ref{fig:comments_issue} graphs the count of comments on an issue
ticket, with the X-axis representing the time a comment was made in
``Anywhere on Earth: AoE'' time. The graph reveals a spike in activity
during office hours (9:00 to 17:00) in AoE time. Similarly, Figure
\ref{fig:comments_PR} presents the count of comments on pull requests,
also indicating a concentration of activity during office hours in the
AoE time zone.

Bias in developers' activity may indicate an uneven distribution of
developers, particularly highly active ones, across different
geographical locations. A 2015 GitHub report revealed that a third of
developers resided in North America. However, countries such as India
and China have seen a significant increase in participation in recent
years \autocite{githubGlobal}.

Alternatively, developers may adhere to AoE time regardless of their
geographical location. For instance, Indian developers working on AoE
office hours would be active from late night to early morning. This
could imply that they are aligning their work schedule with different
time zones to minimize the impact of time differences.

\subsection*{Insight:}

Even within the OSS community, known for its asynchronous activities,
synchronous tasks are still essential in software development. However,
in today's world, where issues like pandemics can arise, it's important
to explore asynchronous software development methods to prevent
companies from halting their work. One possible approach is to implement
a human-in-the-loop system using generative AI-based agents, which could
potentially ease the load on developers.

\printbibliography[segment=\therefsegment,heading=subbibliography]

\clearpage

%% file: English/chapter3_6.tex
\section{OSS community quickly halts their development.}

\subsection*{Question: How much of the OSS community will continue?}

It is true that popular open-source web and database servers have been
continuously developed by the OSS community for many years.The
traditional software development managers object that in software
development it is really sustained effort over time and the degree to
which customers can expect continuing investment in the product that
matters, but the casualness with which project groups form and change
and dissolve in the open-source.\autocite{raymond} Similar to the
situation on GitHub, where over 85.7 million new repositories have been
created and their number continues to increase by 20\%, OSS communities
are being created at a rapid pace\autocite{github2021}. However, it is
important to consider how many of these communities will eventually
cease their activities on a yearly basis. In order to verify this myth,
we will analyze the likelihood of OSS communities being able to sustain
their activities in the long term.

\subsection*{Fact: Four years after adoption, half of the OSS community activities are still ongoing.}

\begin{figure}[hbtp]
\centering
\includegraphics[width=0.9\textwidth,height=0.4\textheight,keepaspectratio]{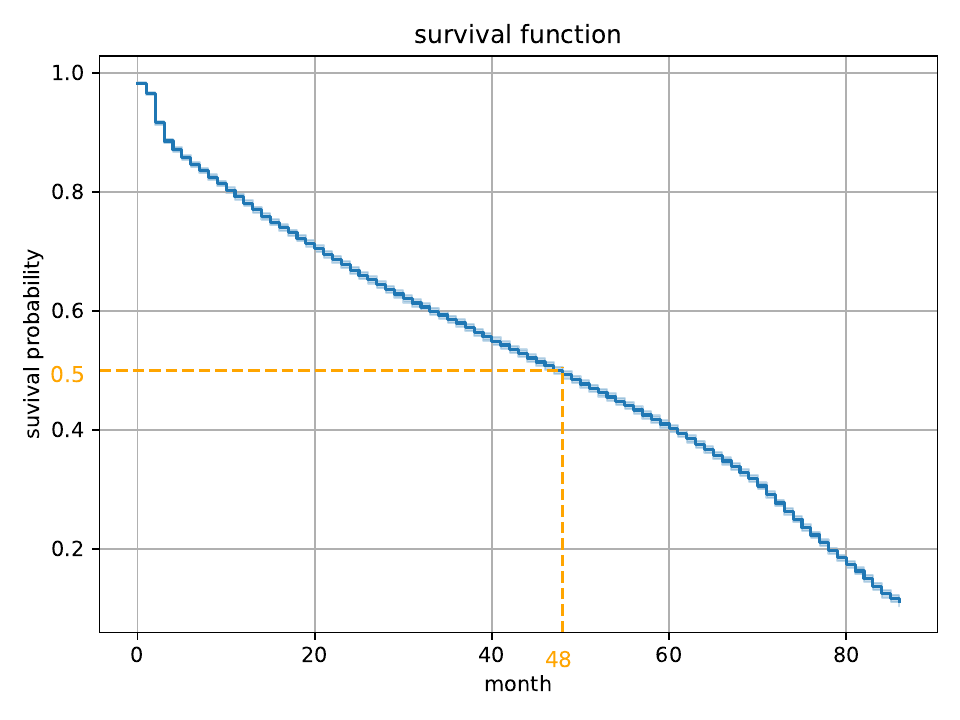}
\caption{The probabilities for the duration of OSS survival time}
\label{fig:OSS_survival_time}
\end{figure}

\begin{figure}[hbtp]
\centering
\includegraphics[width=0.9\textwidth,height=0.4\textheight,keepaspectratio]{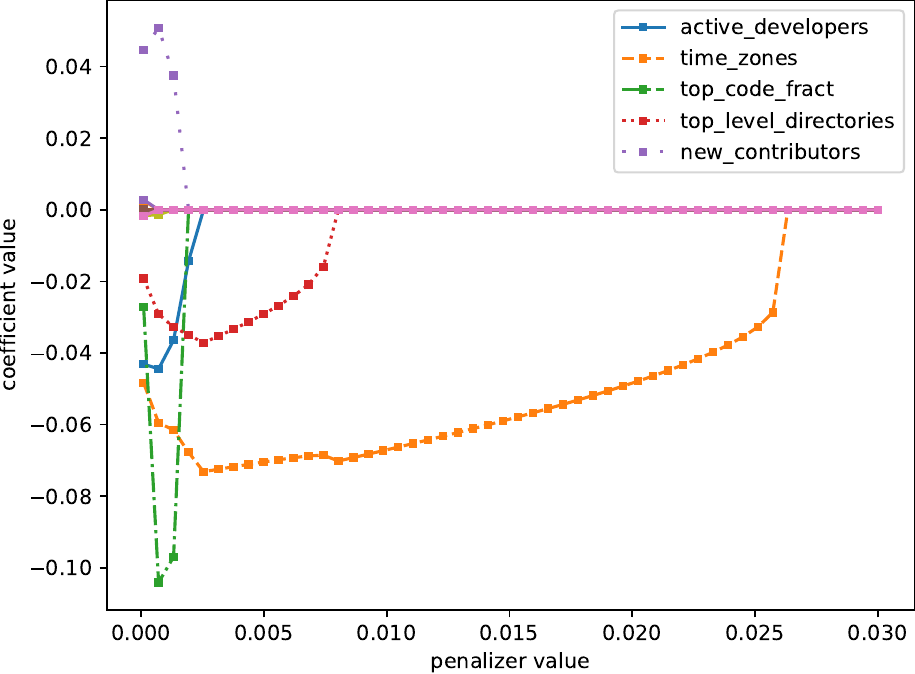}
\caption{The impact of different features on the survival or cessation}
\label{fig:impact_features_survival}
\end{figure}

Figure \ref{fig:OSS_survival_time} illustrates the probabilities for the
duration of OSS survival time. It demonstrates that the probability of
OSSs ceasing development is higher during the first three months
compared to other periods, and the overall trend shows a linear
decrease. The duration until half of the OSSs stop their development is
48 months. Figure \ref{fig:impact_features_survival} illustrates the
impact of different features on the survival or cessation of OSS
development. The vertical axis represents the degree of influence, with
negative values indicating continued development and positive values
indicating cessation. The horizontal axis represents the importance of
the feature, with less important values approaching zero earlier.
According to Figure \ref{fig:impact_features_survival} , the number of
time zones in which developers reside and the number of top-level
directories in OSS have a significant influence on the continuity of OSS
activities.

The decrease in the probability during the first three months is more
severe compared to other intervals. This is because a certain percentage
of the currently active OSS projects are newly created, and many of them
quickly cease their development.

The finding that the diversity of time zones in which developers live
has the greatest impact suggests that diverse OSS communities contribute
to the longevity of OSS projects. Additionally, the number of top-level
directories may also indicate that larger OSS projects tend to continue
their development for a longer period of time in the future.

\subsection*{Insight:}

Long-lived OSS is often cherished by a wide range of individuals. Data
further supports the notion that long-lived OSS attracts developers from
diverse residential areas. This suggests a potential correlation between
participant diversity and the longevity of OSS projects. In the context
of corporate development, enhancing the diversity of individuals engaged
in the development process fosters knowledge sharing among developers
and enhances project quality by incorporating varied perspectives.
Ultimately, this can result in the creation of products that resonate
with a larger audience.

\printbibliography[segment=\therefsegment,heading=subbibliography]

\clearpage

%% file: English/chapter3_14.tex
\section{OSS community does not lose to crackers.}

\subsection*{Question: Does the OSS community take less time to resolve vulnerabilities (security holes)?}

The Cathedral and the Bazaar states Linus's law, which claims that
``Given enough eyeballs, all bugs are shallow.'' This means that with a
sufficient number of developers, all bugs will be found and fixed
promptly, ensuring high software quality \autocite{raymond}. Since OSS
source code is publicly available and the OSS community includes
developers from all over the world, one might expect that all bugs in
OSS will be discovered and resolved immediately. To verify this myth, we
focused on several OSS projects and addressed vulnerabilities that are
known to be more time-consuming to fix compared to common bugs. Our
analysis aimed to determine the time required to fix these
vulnerabilities.

\subsection*{Fact: A vulnerability resolution time of 3 months is not necessarily short.}

\begin{figure}[hbtp]
\centering
\includegraphics[width=0.9\textwidth,height=0.4\textheight,keepaspectratio]{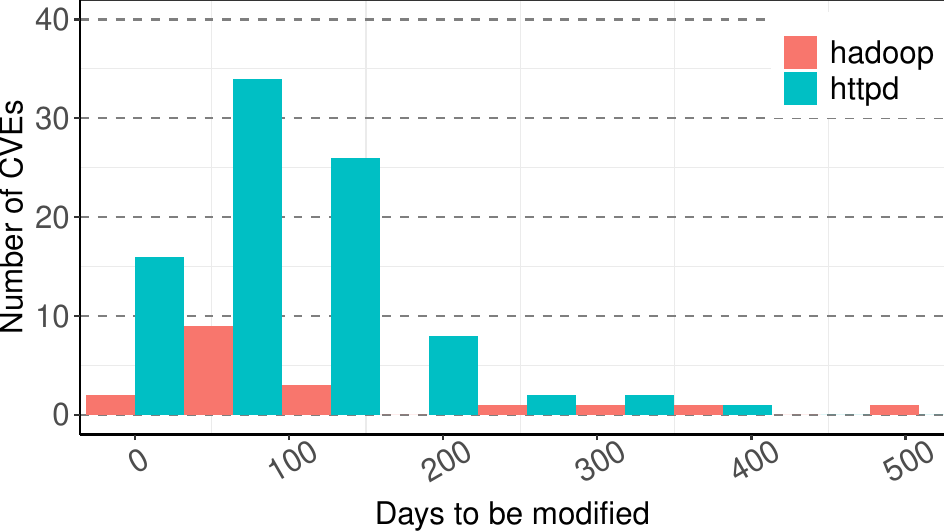}
\caption{The response time to vulnerabilities in Hadoop and httpd}
\label{fig:response_vulnerabilities}
\end{figure}

We analyzed the response time to vulnerabilities in Hadoop and httpd,
two Apache projects with the largest number of developers in OSS. Figure
\ref{fig:response_vulnerabilities} presents the results, with the x-axis
indicating the number of days it takes to fix the vulnerabilities, and
the y-axis indicating the number of vulnerabilities fixed. The figure
shows that vulnerabilities are typically resolved within 100 days. The
median time for both projects is approximately 87 days, which is less
than 3 months.

Although the x-axis in the figure has an upper limit of 500 days due to
its size, in reality, some vulnerabilities took longer than this limit
to be fixed. For instance, the httpd vulnerability that took the longest
time to fix required 1,842 days, which is slightly over five years.

Even in projects with a large number of developers, the fact that it
takes approximately three months to resolve vulnerabilities that
significantly impact quality indicates that it may take time to fix bugs
in OSS. It is worth noting that, although this is an exceptional case,
there are vulnerabilities that have remained unresolved for over five
years. For difficult-to-fix bugs, such as vulnerabilities, the
involvement of multiple individuals may not necessarily speed up the
resolution time.

The analysis conducted so far does not consider the time it takes to
discover a bug. It focuses on the time elapsed after a bug has been
discovered and reported. A previous study reported a median time of 200
days from bug introduction to bug fix {[}1{]}, and our findings suggest
that it may take approximately 100 days to discover a bug.

\subsection*{Insight:}

In certain situations, companies keep developers who are familiar with
the code in order to make post-release code modifications. The speed at
which vulnerabilities are fixed in the OSS community is not particularly
rapid. Nonetheless, the OSS community experiences a high turnover of
developers and manages to address vulnerabilities, even when the
developer who introduced the vulnerability is different from the one who
fixes it. Companies can adopt from the OSS community practices for
handling vulnerabilities in a collaborative manner, rather than relying
solely on individuals.

\printbibliography[segment=\therefsegment,heading=subbibliography]

\clearpage

%% file: English/chapter3_9.tex
\section{OSS community responds quickly to requests.}

\subsection*{Question: Is the resolution time for proposals, like bug reports and enhancements, short?}

The Cathedral and the Bazaar state that ``Treating your users as
co-developers is your least-hassle route to rapid code improvement and
effective debugging.'' In OSS communities, anyone can report bugs and
contribute to code fixes. Early and frequent releases are a critical
part of the OSS development model. \autocite{raymond}. These factors
have led some to assume that any request for open-source software (OSS)
will be addressed immediately. To verify this myth, we will examine the
resolution time for two types of proposals: bug reports and feature
requests.

\subsection*{Fact: Most bug reports and feature requests are resolved within two weeks. However, at least a quarter of them take more than three months to resolve, indicating a variation in resolution time.}

\begin{figure}[hbtp]
\centering
\includegraphics[width=0.9\textwidth,height=0.4\textheight,keepaspectratio]{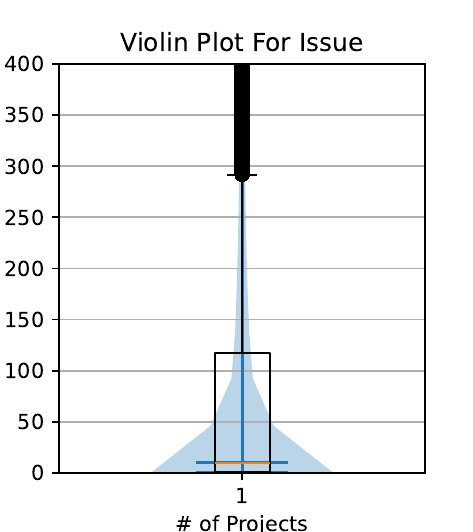}
\caption{The response times in bug reports}
\label{fig:response_bug}
\end{figure}

\begin{figure}[hbtp]
\centering
\includegraphics[width=0.9\textwidth,height=0.4\textheight,keepaspectratio]{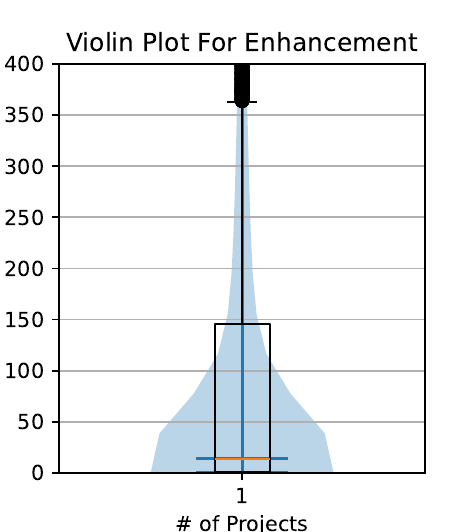}
\caption{The response times in feature requests}
\label{fig:response_request}
\end{figure}

Figure \ref{fig:response_bug} illustrates the boxplot and violin plot
for response times in bug reports. The median response time is 10 days,
with the 75th percentile at 117 days. Figure \ref{fig:response_request}
presents the boxplot and violin plot for response times in feature
requests, with the median and 75th percentile at 14 and 146 days
respectively. Half of both bug reports and feature requests are resolved
within two weeks. However, 25\% of bug reports and feature requests take
more than three and four months to resolve respectively.

Although approximately half of bug reports and feature requests are
resolved within two weeks, it takes more than four months to resolve
three-quarters of them. This finding suggests that not all users who
submit bug reports or feature requests actively contribute to
development, and there may be a limited number of developers compared to
the volume of incoming reports. Additionally, the effort required to
resolve these reports and requests varies greatly. Some are closed
without much discussion, while others require extensive deliberation
before being resolved. With limited resources (developers), there is a
constraint on the number of requests that can be addressed, making
request triage crucial.

\subsection*{Insight:}

Even in OSS development, which can gather developers from all over the
world, it is challenging to immediately address all bug reports and
feature requests. The findings of this study highlight this reality.
Proper triage of requirements is crucial to maximize limited response
capabilities. By appropriately triaging requirements, developers can
manage their workload within a suitable range. Based on research on
triage methods in OSS development, implementing these methods will
optimize development in the enterprise.

\printbibliography[segment=\therefsegment,heading=subbibliography]

\clearpage

%% file: English/chapter3_5.tex
\section{OSS community participants are top-notch developers.}

\subsection*{Question: What roles do developers play in the OSS Community?}

The OSS community is full of famous hackers. In recent years, developers
and corporate recruiters have recognized that the talent of OSS
developers often leads to employment and high salaries
\autocite{edx10th}. Consequently, there is a prevailing belief that
engineers who actively participate in OSS communities are exceptionally
talented and regarded as top-notch developers. As of January 2023, there
are more than 100 million engineers using GitHub
\autocite{github100million}, and the OSS community is still growing. If
the above belief is correct, does this large-sized OSS community also
consist of top-notch developers? To verify this myth, we analyze the
roles of developers participating in OSS communities.

\subsection*{Fact: In the OSS community, there is a wide range of roles available.}

\begin{figure}[hbtp]
\centering
\includegraphics[width=0.9\textwidth,height=0.4\textheight,keepaspectratio]{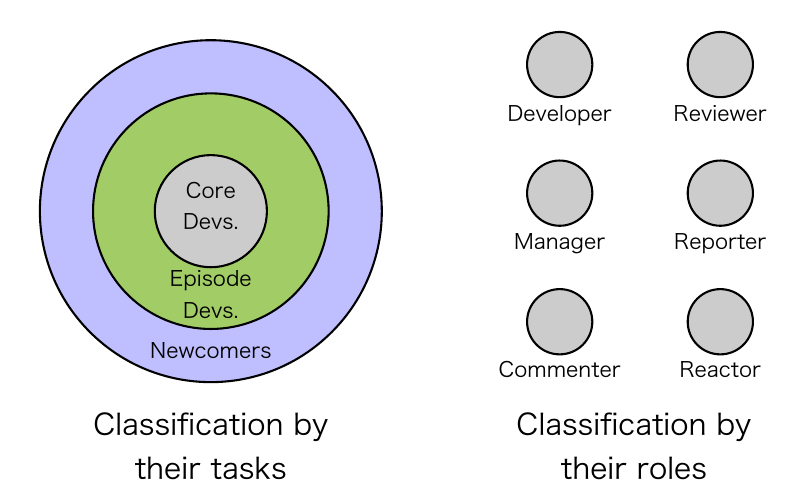}
\caption{The roles in OSS / corporate development}
\label{fig:role_OSS}
\end{figure}

We conducted a survey of previous studies on the role of developers in
OSS communities and the structure of these communities. In most of the
previous studies, the existence of a core team has been assumed or
observed as the structure of the communities \autocite{roblesMsr09}.
Although the classification of developer roles varies from study to
study, it has been assumed and observed that there are core developers
who play a central role, episodic developers who participate on a
limited basis by adding specific features, and new developers who are
new to the community \autocite{roblesMsr09},\autocite{barcombTse20}.
Additionally, there are other roles in OSS that differ from those in
corporate development, such as reactors \autocite{yueTse23}, who
primarily provide reactive feedback (Figure \ref{fig:role_OSS}).

Some OSS communities have implemented strategies to facilitate the
involvement of developers with limited experience but a strong desire to
participate. They label certain tasks, such as documentation updates and
minor bug fixes, as ``Good First Issues'' to make them more accessible
to newcomers. In addition, reactors in the community can provide
valuable feedback regardless of their technical expertise, as long as
their focus is not on designing new features, but rather on determining
the desirability of specific features. It is important to note that the
roles of developers in the OSS community are diverse, and participation
in OSS development does not guarantee their technical skills.

\subsection*{Insight:}

There was a time when companies focused on teams made up of only the
best developers, such as Green Beret talent. On the other hand,
developers participating in OSS development have a wide variety of
skills and ways to contribute. We found that there are roles that are
not considered important by companies, such as not only coding and
documentation, but also responding to discussions. The diversity of
roles in OSS may allow companies to learn new values.

\printbibliography[segment=\therefsegment,heading=subbibliography]

\clearpage

%% file: English/chapter4.tex
\chapter{Takeaway messages}

In Chapter 3, we discussed the six myths and evaluated their validity,
as well as the insights they provide. In this chapter, we will summarize
these myths from different perspectives and explore what can be learned
from the development style of the OSS community. Additionally, we will
address the key issues that need to be tackled in order to achieve
Employee Experience (EX) in software development within companies and
organizations.

Our main goal is to create more opportunities for individuals from
diverse backgrounds to succeed in software development. We aim to make
software development accessible to anyone, at any time, and from
anywhere. Developers from all over the world are actively participating
in the OSS community. However, there are still a few challenges that
need to be addressed.

\section{Does OSS allow everyone to participate in the community?}

An actual situation in OSS: Based on the validation that ``OSS community
participants are top-notch developers,'' it appears that the
contribution role is divided into the core team and others. While not
all OSS communities are making efforts to reduce the cost of
participation, some are actively working towards this goal. It is
important to note that not everyone is involved, but steps are being
taken to achieve greater inclusivity.

Challenge: The knowledge and skills required for software development
are becoming more diverse. Instead of specializing in just one field,
engineers with a wide range of knowledge and skills, such as full-stack
engineers, are in high demand. The journey to becoming a developer and
participating in projects is getting longer and longer.

Takeaway message: To incorporate OSS development styles, it is important
to identify practices that allow every participant to contribute to the
project using their knowledge and skills. It is also crucial to find
ways to improve practical skills through active participation in the
project. Analyzing how actions other than coding and testing (e.g.,
reactions) impact software and development projects can uncover new
insights and values.

\section{Is the OSS anytime/anywhere available to the community?}

An actual situation in OSS: Based on the validation that ``OSS community
quickly halts their development,'' we have observed that the
participants' various time zones may contribute to the project's
continuous progress. The OSS community consists of developers from all
over the world. Based on the validation that ``OSS community never
sleeps'' and ``Communication within the OSS is moderate,'' the activity
time is concentrated within a specific time period, and it is observed
that the communication was synchronous with short intervals. It appears
that ``anywhere'' is generally achievable through the network, but the
same cannot be said for ``anytime''.

Challenge: In software development projects, it is common practice to
keep developers in close proximity to facilitate communication.
Companies often assign fixed work locations for developers to manage
confidential information found in source code and designs. While
networks and tools have reduced the limitations of physical locations,
they still persist. Furthermore, if a developer is unable to participate
in communication during certain times of the day, they may be unable to
fully engage in a software development project.

Takeaway message: To overcome communication delays and increase
geographical freedom, it will be necessary to establish a new
development style and tools that prioritize information management.

\section{Product management in OSS}

An actual situation in OSS: Based on the validation that ``OSS community
responds quickly to requests'' and ``OSS community does not lose to
crackers,'' We have discovered that not all requests and bugs are
resolved promptly in OSS.

Challenge: Even in companies, bugs and requests are coming in large
numbers, not only from traditional channels but also from various
sources such as social media and reviews. There are insufficient
developers to address all of them.

Takeaway message: To effectively manage a large number of requirements
and bug reports, it is important to learn how to triage them based on
their priority and urgency. This involves identifying which issues need
to be addressed first. Additionally, it may be beneficial to develop
tools specifically designed for this purpose. In corporate software
development, it can be effective to consider approaches like inner
sourcing to encourage diverse contributions to non-critical requests and
bugs.

%% file: common/colphan.tex
\clearpage
\thispagestyle{empty}

\vspace*{\fill}
\begin{flushleft}
    \begin{tabular*}{\textwidth}{@{}l@{\extracolsep{\fill}}}
        \textbf{\huge OSS Myths and Facts} \\
        \hline
        \begin{tabular}{@{}r@{/}r@{/}r}
            2024 & 04 & 1 \\
        \end{tabular} \\
        \\
        \begin{tabular}{@{}l}
            \CRtext
        \end{tabular} \\
        \hline
    \end{tabular*}
\end{flushleft}

%% file: OSS_myths_and_facts.bib
@book{cockburn,
    author    = {Cockburn Alistair},
    title     = {Agile software development: the cooperative game.},
    publisher = {Pearson Education},
    year      = {2006},
}

@book{raymond,
    author    = {Raymond Eric},
    title     = {The cathedral and the bazaar.},
    publisher = {O'Reilly Media, Inc.},
    year      = {2001},
}

@inproceedings{kondoCompsac23,
    author={M. Kondo, S. Saito, and Y. Iimura and Choi E. and O. Mizuno, and Y. Kamei and N. Ubayashi},
    booktitle={2023 IEEE 47th Annual Computers, Software, and Applications Conference (COMPSAC)}, 
    title={Towards Better Online Communication for Future Software Development in Industry}, 
    year={2023},
    volume={},
    number={},
    pages={1619-1624},
    doi={10.1109/COMPSAC57700.2023.00250}
}

@article{roblesMsr09,
    author={Robles Gregorio and Gonzalez-Barahona Jesus M. and Herraiz Israel},
    title={Evolution of the core team of developers in libre software projects}, 
    booktitle={MSR '09: Proceedings of the 2009 6th IEEE International Working Conference on Mining Software Repositories},
    year={2009},
    pages={167-170},
    number={},
    volume={},
    publisher = {IEEE},
    doi={10.1109/MSR.2009.5069497}
}

@article{barcombTse20,
    author={Barcomb Ann and Kaufmann Andreas and Riehle Dirk and Stol Klaas-Jan and Fitzgerald Brian},
    journal={IEEE Transactions on Software Engineering}, 
    title={Uncovering the Periphery: A Qualitative Survey of Episodic Volunteering in Free/Libre and Open Source Software Communities}, 
    year={2020},
    volume={46},
    number={9},
    pages={962-980},
    doi={10.1109/TSE.2018.2872713}
}

@article{yueTse23,
    author={Yue Yang and Wang Yi and Redmiles David},
    journal={IEEE Transactions on Software Engineering}, 
    title={Off to a Good Start: Dynamic Contribution Patterns and Technical Success in an OSS Newcomer’s Early Career}, 
    year={2023},
    volume={49},
    number={2},
    pages={529-548},
    doi={10.1109/TSE.2022.3156071}
}

@misc{githubGlobal,
    author        = {GitHub},
    title         = {A global community of developers},
    howpublihshed = {\url{https://octoverse.github.com/2022/global-tech-talent}},
    note          = {(Accessed on December 1, 2023)}
}

@misc{github2021,
    author        = {GitHub},
    title         = {The 2021 State of the Octoverse},
    howpublihshed = {\url{https://octoverse.github.com/2021/}},
    note          = {(Accessed on December 1, 2023)}
}

@misc{github100million,
    author        = {GitHub},
    title         = {100 million developers and counting},
    howpublihshed = {\url{https://github.blog/2023-01-25-100-million-developers-and-counting/}},
    note          = {(Accessed on December 1, 2023)}
}

@misc{edx10th,
    author        = {edX and Linux Foundation Research},
    title         = {10th Annual Open Source Jobs Report},
    howpublihshed = {\url{https://training.linuxfoundation.org/resources/2022-open-source-jobs-report/}},
    note          = {(Accessed on December 1, 2023)}
}

@misc{microsoft22,
    author        = {Teevan, Jaime and Baym, Nancy and Butler et al},
    full_author = {Teevan, Jaime and Baym, Nancy and Butler, Jenna and Hecht, Brent and Jaffe, Sonia and Nowak, Kate and Sellen, Abigail and Yang, Longqi and Ash, Marcus and Awori, Kagonya and Bruch, Mia and Choudhury, Piali and Coleman, Adam and Counts, Scott and Cupala, Shiraz and Czerwinski, Mary and Doran, Ed and Fetterolf, Elizabeth and Gonzalez Franco, Mar and Gupta, Kunal and Halfaker, Aaron L and Hadley, Constance and Houck, Brian and Inkpen, Kori and Iqbal, Shamsi and Knudsen, Eric and Levine, Stacey and Lindley, Siân and Neville, Jennifer and O’Neill, Jacki and Pollak, Rick and Poznanski, Victor and Rintel, Sean and Shah, Neha Parikh and Suri, Siddharth and Troy, Adam D. and Wan, Mengting},
    title         = {Microsoft New Future of Work Report 2022},
    howpublihshed = {\url{https://www.microsoft.com/en-us/research/publication/microsoft-new-future-of-work-report-2022/}},
    note          = {(Accessed on December 1, 2023)}
}
